\documentclass[]{article}
\usepackage{amsmath}
\usepackage{threeparttablex}
\usepackage{booktabs}

\title{Approximating Bayes factors from minimal ANOVA summaries: An extension of the BIC method}
\author{Thomas J. Faulkenberry \thanks{Email: \texttt{faulkenberry@tarleton.edu}}\\
  Department of Psychological Sciences\\
  Tarleton State University, USA
}

\begin{document}
\maketitle

The purpose of this brief report is to extend a method of Masson \cite{masson2011} to an easy-to-use formula for performing Bayesian hypothesis tests from minimal ANOVA summaries.  The original method, which was based on earlier work by Raftery \cite{raftery1995} and Wagenmakers \cite{wagenmakers2007}, uses the BIC (Bayesian Information Criterion) to compute an estimate of the Bayes factor $BF_{01}$, a ratio which indexes the extent to which incoming data updates the prior odds in favor of the null hypothesis over the alternative hypothesis.  Unfortunately, the method of Masson \cite{masson2011} requires access to the sum-of-squares values from the original ANOVA.  This is not a problem for the researcher who is analyzing raw data, but if one only has access to the published ANOVA summary, the computation is nontrivial without access to the sum-of-squares values.  The method described here requires only knowledge of the number of subjects, degrees of freedom, and the $F$ statistic, all of which are usually provided in even the most minimal ANOVA summaries.

\section{The BIC approximation of the Bayes factor}

Based on work by Raftery \cite{raftery1995}, Wagenmakers \cite{wagenmakers2007} demonstrated a method for estimating Bayes factors using the BIC. For a given model $H_i$, the BIC is defined as

\[
  \text{BIC}(H_i) = -2\log L_i + k_i\cdot \log n,
\]

\noindent
where $n$ is the number of observations, $k_i$ is the number of free parameters
of model $H_i$, and $L_i$ is the maximum likelihood for model $H_i$.  He then showed that the Bayes factor for $H_o$ over $H_1$ can be approximated as

\begin{equation}
  \label{eq:BF}
  BF_{01} \approx \exp\Bigl(\Delta\text{BIC}_{10}/2\Bigl),
\end{equation}

\noindent
where $\Delta\text{BIC}_{10} = \text{BIC}(H_1)-\text{BIC}(H_0)$.  Further, Wagenmakers \cite{wagenmakers2007} showed that when comparing an alternative hypothesis $H_1$ to a null hypothesis $H_0$, 

\begin{equation}\label{eq:BIC}
  \Delta\text{BIC}_{10} = n\log\Biggl(\frac{SSE_1}{SSE_0}\Biggr) + (k_1-k_0)\log n. 
\end{equation}

\noindent
In this equation, $SSE_0$ and $SSE_1$ represent the sum of squares for the error terms in models $H_0$ and $H_1$, respectively.  Both Wagenmakers \cite{wagenmakers2007} and Masson \cite{masson2011} give excellent examples of how to use this approximation to compute Bayes factors, assuming one is given information about $SSE_0$ and $SSE_1$, as is the case with most statistical software.  However, if one is only given the ANOVA summary (e.g., $F(1,23)=4.35$), the computation is nontrivial.  This motivates my development below of a formula which does not require the sum-of-squares values.

To begin, suppose we wish to examine an effect of some independent variable with associated $F$-ratio $F(df_1,df_2)$, where $df_1$ represents the degrees of freedom associated with the manipulation, and $df_2$ represents the degrees of freedom associated with the error term.  Then,

\[
  F = \frac{SS_1/df_1}{SS_2/df_2} = \frac{SS_1}{SS_2}\cdot \frac{df_2}{df_1},
\]

\noindent
where $SS_1$ and $SS_2$ are the sum of squared errors associated with the manipulation and the error term, respectively. 

From Equation \ref{eq:BIC}, we see that

\begin{align*}
  \Delta\text{BIC}_{10} &= n\log\left(\frac{SSE_1}{SSE_0}\right) + (k_1-k_0)\log n\\
                        & = n\log \left(\frac{SS_2}{SS_1+SS_2}\right) + df_1\log n.
\end{align*}

\noindent
This equality holds because $SSE_1$ represents the sum of squares that is not explained by $H_1$, which is simply $SS_2$ (the error term).  Similarly, $SSE_0$ is the sum of squares not explained by $H_0$, which is the sum of $SS_1$ and $SS_2$ (see \cite{wagenmakers2007}, page 799).  Finally, in the context of comparing $H_1$ and $H_0$ in an ANOVA design, we have $k_1-k_0=df_1$.  Now, we can use algebra to re-express $\Delta\text{BIC}_{10}$ in terms of $F$:

\begin{align*}
  \Delta\text{BIC}_{10} &=  n\log \left(\frac{SS_2}{SS_1+SS_2}\right) + df_1\log n\\
  &= n\log\left(\frac{1}{\frac{SS_1}{SS_2}+1}\right) + df_1\log n\\
                        &= n\log \left( \frac{\frac{df_2}{df_1}}{\frac{SS_1}{SS_2}\cdot \frac{df_2}{df_1}+\frac{df_2}{df_1}}\right) + df_1\log n\\
                        &= n\log \left(\frac{\frac{df_2}{df_1}}{F + \frac{df_2}{df_1}}\right) + df_1\log n\\
  & = n\log\left(\frac{df_2}{Fdf_1 + df_2}\right) + df_1\log n.\\
\end{align*}

\noindent
Substituting this into Equation \ref{eq:BF}, we can compute:
\begin{align*}
  BF_{01} & = \exp\left(\Delta\text{BIC}_{10}/2\right)\\
  & = \exp\left[ \frac{1}{2} \left( n\log\left(\frac{df_2}{Fdf_1 + df_2}\right) + df_1\log n\right)\right]\\
          & = \exp \left[\frac{n}{2}\log\left(\frac{df_2}{Fdf_1+df_2}\right) + \frac{df_1}{2}\log n\right]\\
          & = \left(\frac{df_2}{Fdf_1+df_2}\right)^{n/2} \cdot n^{df_1/2}\\
          & = \sqrt{\frac{df_2^n \cdot n^{df_1}}{(Fdf_1+df_2)^n}}\\
          &= \sqrt{\frac{n^{df_1}}{\left(\frac{Fdf_1}{df_2}+1\right)^n}}.
\end{align*}

Rearranging this last expression slightly yields the equation:
\begin{equation}\label{eq:BIC2}
BF_{01} = \sqrt{n^{df_1}\left(1+\frac{Fdf_1}{df_2}\right)^{-n}}
\end{equation}

\noindent
Note that since $BF_{10} = 1/BF_{01}$, the formula can be used flexibly to assess evidence for either the null hypothesis or the alternative hypothesis, depending on the researcher's needs.

\subsection{An example}

Fayol and Thevenot \cite{fayol2012} used an operator priming paradigm to study the processes involved in mental addition.  Specifically, they recorded vocal onset latencies (in milliseconds) as 18 participants answered addition and multiplication problems in two conditions.  One condition presented the operation sign ($+$ or $\times$) 150 msec before the numerical operands.  The other condition presented the operation sign simultaneously with the operands.  While Fayol and Thevenot found that this 150 msec preview resulted in a significant speed-up for addition problems, they reported no such speed-up for multiplication, $F(1,17)=1.75$, $p=0.20$.  This was claimed as evidence that addition and multiplication are subject to different mental processes (i.e., multiplication is memory-based whereas addition is procedural).

However, using a nonsignificant result in traditional hypothesis testing as ``evidence'' for a null effect is problematic \cite{wagenmakers2007}.  To mitigate this problem, one can directly compute evidence for the null with a Bayes factor $BF_{01}$.  Equation \ref{eq:BIC2} gives us

\begin{align*}
BF_{01} &= \sqrt{n^{df_1}\left(1+\frac{Fdf_1}{df_2}\right)^{-n}}\\
        & = \sqrt{18^1\left(1+\frac{1.75\cdot 1}{17}\right)^{-18}}\\
  &= 1.757.
\end{align*}

\noindent
Hence, we see that, at best, the data only yields weak evidence in favor of the null hypothesis.

\section{Simulations: BIC approximation versus Bayesian default ANOVA}

At this stage, it is clear that Equation \ref{eq:BIC2} provides a straightforward method for computing an approximate Bayes factor given only minimal output from a reported ANOVA.  However, it is not yet clear to what extent this BIC approximation would result in the same decision if a Bayesian analysis of variance \cite{rouder2012} were performed on the raw data.  To answer this question, I performed a series of simulations.

Each simulation consisted of 1000 randomly generated data sets under a $2\times 3$ factorial design.  I simulated 3 different cell-size conditions: $n=20, 50,$ or $80$. Specifically each data set consisted of a vector $\mathbf{y}$ generated as

\[
  y_{ijk} = \alpha_i + \tau_j + \gamma_{ij} +\varepsilon_{ijk}
\]

\noindent
where $i=1,2$, $j=1,2,3$, and $k=1,\dots, n$.  The ``effects'' $\alpha$, $\tau$, and $\gamma$ were generated from multivariate normal distributions with mean 0 and variance $g$, yielding three different effect sizes obtained by setting $g = 0, 0.05,$ and $0.2$ \cite{wang2017}.  In all, there were 9 different simulations, generated by crossing the 3 cell sizes ($n=20, 50, 80$) with the 3 effect sizes ($g=0, 0.05, 0.2$).

For each data set, I computed (1) a Bayesian ANOVA using the BayesFactor package in R \cite{bayesfactor} and (2) the BIC approximation using Equation \ref{eq:BIC2} from the traditional ANOVA.  Bayes factors were computed as $BF_{10}$ to assess evidence in favor of the alternative hypothesis over the null hypothesis.  Similar to \cite{wang2017}, I set the decision criterion to select the alternative hypothesis if $\log(BF)>0$, and the null hypothesis otherwise.  Five-number summaries for $\log(BF)$ are reported for the $n=50$ simulation in Table \ref{tab:summary}, as well as the proportion of simulated data sets for which the Bayesian ANOVA and the BIC approximation from Equation \ref{eq:BIC2} selected the same model.  For brevity, I only report the results for the interaction $\gamma$; similar results hold for the main effects $\alpha$ and $\tau$, as well as for the $n=20$ and $n=50$ conditions.

\begin{table*}[h]
	\vspace*{2em}
	\begin{threeparttable}
	\caption{Summary of simulation results for $n=50$}
	\label{tab:summary}
	\begin{tabular}{cccccccc} 
          \toprule
          $g$ & $BF$ type & Min & $Q_1$ & Median & $Q_3$ & Max & Consistency\\
          \midrule
                    0 & BayesFactor & -4.88 & -3.08 & -2.70 & -2.10 & 3.93 & \\
                    & BIC & -3.40 & -3.12 & -2.70 & -2.04 & 4.60 & 0.991 \\[2mm]

                    0.05 & BayesFactor & -3.40 & -2.39 & -1.18 & 1.02 & 28.38 & \\
              & BIC & -3.40 & -2.35 & -0.99 & 1.41 & 30.13 & 0.968\\[2mm]

          0.2 & BayesFactor & -3.37 & -0.45 & 3.03 & 10.06 & 52.04 & \\
              & BIC & -3.40 & -0.19 & 3.62 & 11.24 & 55.32 & 0.976\\
          \midrule

	\end{tabular}
        \begin{tablenotes}
        \item {\small\textit{Note:} All BayesFactor models were fit with a ``wide'' prior, which is roughly equivalent to the unit-information prior used by Raftery \cite{raftery1995} for the BIC approximation.}
        \end{tablenotes}

	\end{threeparttable}
\end{table*}

As shown in Table \ref{tab:summary}, the BIC approximation from Equation \ref{eq:BIC2} provides a similar distribution of Bayes factors compared to those computed from the BayesFactor package in R.  Additionally, the two resulted in the same decision in a large proportion of simulations.

\section{Conclusion}
The BIC approximation given in Equation \ref{eq:BIC2} provides an easy-to-use estimate of Bayes factors for ANOVA designs.  It requires only minimal information, which makes it well-suited for using in a meta-analytic context.  In simulations, the estimates derived from Equation \ref{eq:BIC2} compare favorably to Bayes factors computed from raw data.  Thus, the researcher can confidently add this BIC approximation to the ever-growing collection of Bayesian tools for psychological measurement.

\bibliographystyle{plain}
\bibliography{references.bib}

\end{document}